\documentclass[superscriptaddress,twocolumn,showpacs,preprintnumbers,amsmath,amssymb,floatfix,nofootinbib]{revtex4-1}

\usepackage{color}   
\usepackage{graphicx}
\usepackage{dcolumn}

\begin{document}

\def\bb    #1{\hbox{\boldmath${#1}$}}

\title{Intriguing properties of multiplicity distributions}

\author{Maciej Rybczy\'nski}
\email{maciej.rybczynski@ujk.edu.pl}
\affiliation{Institute of Physics, Jan Kochanowski University, 25-406 Kielce, Poland}
\author{Grzegorz Wilk}
\email{grzegorz.wilk@ncbj.gov.pl}
\affiliation{ National Centre for Nuclear Research, Warsaw 00-681, Poland}
\author{Zbigniew W\l odarczyk}
\email{zbigniew.wlodarczyk@ujk.edu.pl}
\affiliation{Institute of Physics, Jan Kochanowski University, 25-406 Kielce, Poland}

\begin{abstract}
Multiplicity distributions exhibit, after closer inspection, peculiarly enhanced void probability
and oscillatory behavior of the modified combinants. We discuss the possible sources of these oscillations
and their impact on our understanding of the multiparticle production mechanism. Theoretical
understanding of both phenomena within the class of compound distributions is presented.
\end{abstract}


\maketitle
\section{Introduction}
\label{Introduction}

Multiplicity distributions, $P(N)$, are among the first observables measured in any multiparticle production experiment and are among the most thoroughly investigated and discussed sources of information on the mechanism of the production process \cite{Kittel}.   Nevertheless, it seems that some of their properties remain unnoticed or unused as a possible source of such information. In this work we analyse the non-single diffractive (NSD) charged multiplicity distributions concentrating on two features: $(i)$ on the observation that, after closer inspection, they show a peculiarly enhanced void probability, $P(0) > P(1)$ \cite{EVP,Void}, and $(ii)$ on the oscillatory behavior of the so called modified combinants, $C_j$, introduced by us in \cite{JPG,IJMPA}. We demonstrate how these modified combinants can be extracted experimentally from the measured $P(N)$ by means of some recurrence relation involving all $P(N<j)$, and argue that they contain information (located mainly in the small $N$ region) which was so far not disclosed and used. This information is hidden in the specific distinct oscillatory behavior of the $C_j$, which, in most cases, is not observed in the $C_j$ obtained from the $P(N)$ commonly used to fit experimental results. We discuss the possible sources of such behavior and the connection of the $C_j$ with the enhancement of void probabilities, and their impact on our understanding of the multiparticle production mechanism with the emphasis on the theoretical understanding of both phenomena within the class of compound distributions.

\vspace*{-0.2cm}

\section{Modified combinants, combinants and void probabilities}
\label{MC-C-Void}

The dynamics of the multiparticle production process is hidden in the way in which the consecutive measured multiplicities $N$ are connected. In the simplest case one assumes that the multiplicity $N$ is directly influenced only by its neighboring multiplicities $(N \pm 1)$ in the way dictated by the simple recurrence relation:
\begin{equation}
(N+1)P(N+1) = g(N)P(N),\quad g(N) = \alpha + \beta N.\label{rr1}
\end{equation}
The most popular forms of $P(N)$ emerging from this recurrence relation are: the Binomial Distribution (BD) (for which $\alpha = Kp/(1-p)$ and $\beta = -\alpha/K)$,
\begin{equation}
P_{BD}(N) = \frac{K!}{N!(K - N)!} p^N (1 - p)^{K-N},\label{BD}
\end{equation}
the Poisson Distribution (PD) (for which $\alpha = \lambda$ and $\beta = 0$),
\begin{equation}
P_{PD}(N) = \frac{\lambda^N}{N!} \exp( - \lambda), \label{PD}
\end{equation}
and the Negative Binomial Distribution (NBD) (for which $\alpha = kp$ and $\beta = \alpha/k$, where $p$ denotes the probability of particle emission),
\begin{equation}
P_{NBD}(N) = \frac{\Gamma(N+k)}{\Gamma(N+1)\Gamma(k)} p^N (1 - p)^k.\label{NBD}
\end{equation}

Usually the first choice of $P(N)$ in fitting data is a single NBD \cite{KNO}. However, with growing energy and  number of produced secondaries the NBD increasingly deviates from data for large $N$ (see \cite{JPG}) and is replaced either by combinations of two \cite{GU,PG}, three \cite{Z}, or multi-component NBDs \cite{DN}, or by some other form of $P(N)$ \cite{Kittel,KNO,DG,MF,HC}. However, such a procedure only improves the agreement at large $N$, whereas the ratio $R = data/fit$ deviates dramatically from unity at small $N$ for all fits \cite{JPG,IJMPA}. This observation, when taken seriously, suggests that there is some additional information in the measured $P(N)$ not covered by the recurrence relation (\ref{rr1}), which is too restrictive. In \cite{JPG} we proposed a more general form of the recurrence relation, used in counting statistics when dealing with multiplication effects in point processes \cite{ST}. Contrary to Eq. (\ref{rr1}), it now connects all multiplicities by means of some coefficients $C_j$, which define the corresponding $P(N)$ in the following way:
\begin{equation}
(N + 1)P(N + 1) = \langle N\rangle \sum^{N}_{j=0} C_j P(N - j). \label{rr2}
\end{equation}
The coefficients $C_j$ contain the memory of particle $N+1$ about all the $N-j$ previously produced particles. They can be directly calculated from the experimentally measured $P(N)$ by reversing Eq. (\ref{rr2}) and putting it in the form of the following recurrence formula for $C_j$ \cite{JPG}:
\begin{equation}
\langle N\rangle C_j = (j+1)\left[ \frac{P(j+1)}{P(0)} \right] - \langle N\rangle \sum^{j-1}_{i=0}C_i \left[ \frac{P(j-i)}{P(0)} \right]. \label{rCj}
\end{equation}

\begin{figure}[h]
\begin{center}
\includegraphics[scale=0.45]{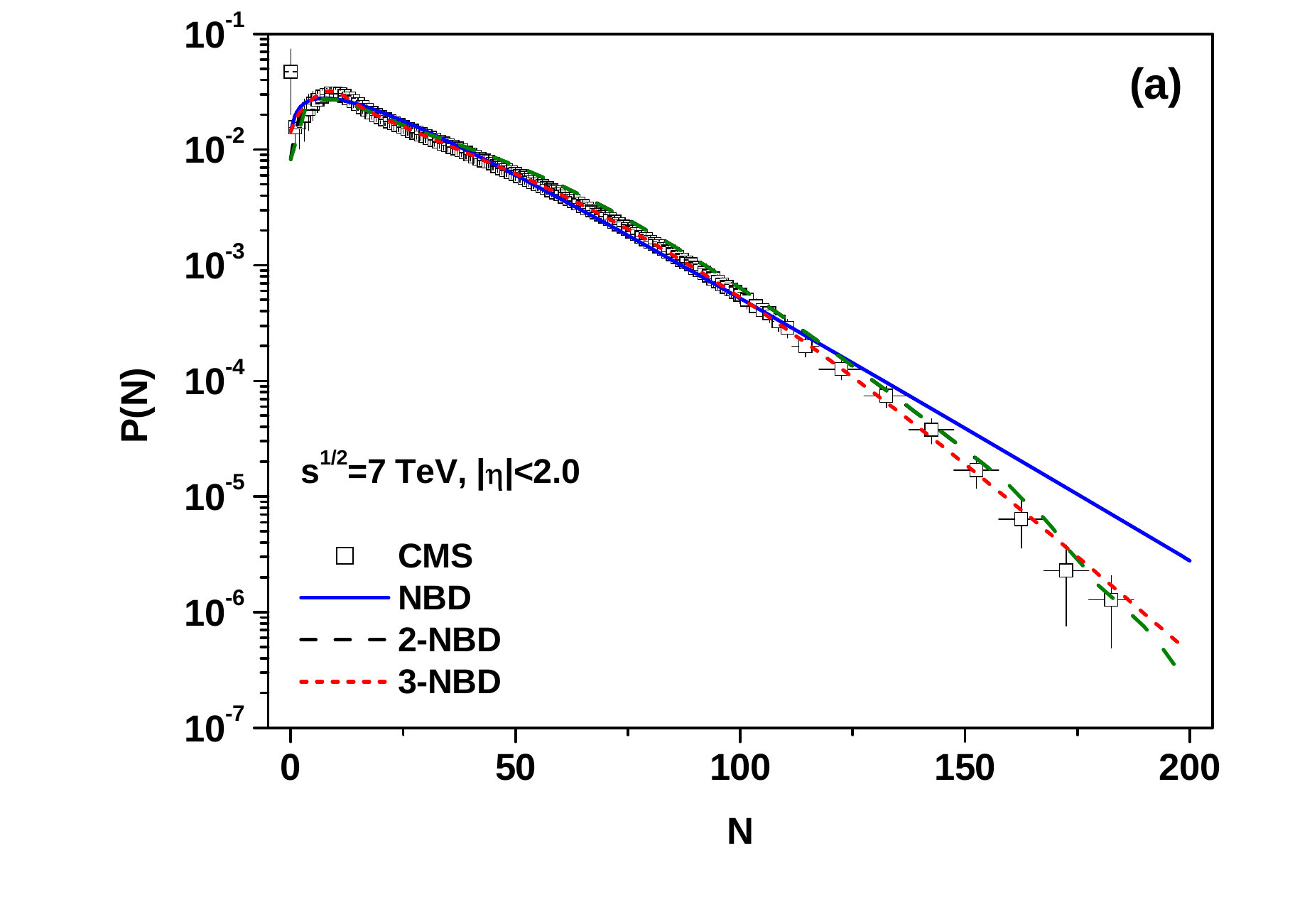}\\
\includegraphics[scale=0.45]{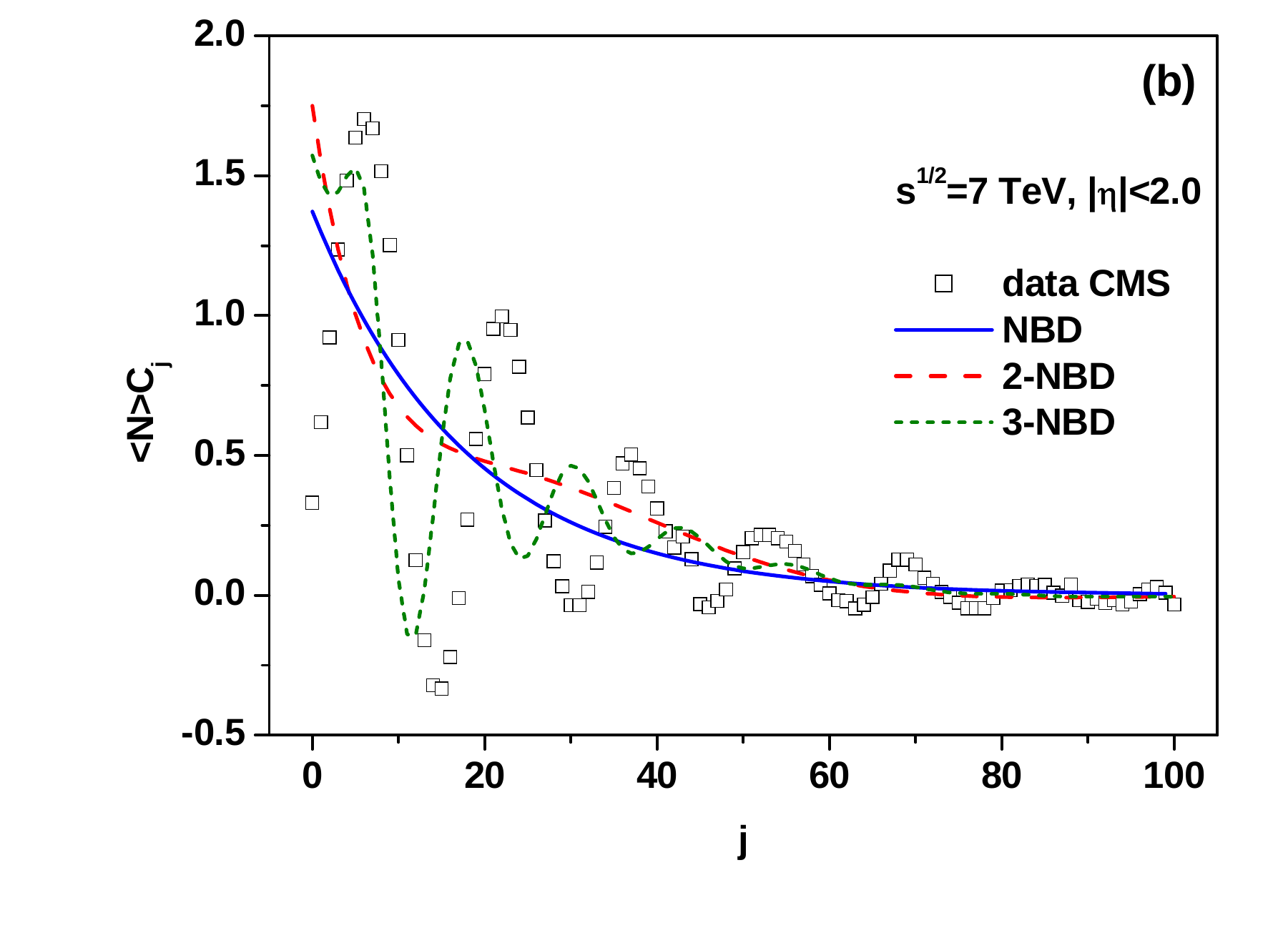}
\end{center}
\vspace{-7mm}
\caption{(Color online) $(a)$ Charged hadron multiplicity distributions for the pseudorapidity range $|\eta| < 2$ at $\sqrt{s} = 7$ TeV, as given by the CMS experiment \cite{CMS} (squares), compared with a NBD for parameters $\langle N\rangle = 25.5$ and $k = 1.45$ (full blue line), with the $2$-component NBD  with parameters from \cite{PG} (red dashed line) and with a $3$-component NBD with parameters from \cite{Z} (dotted green line). $(b)$ The corresponding modified combinants $C_j$ emerging from the CMS data (squares) compared with the same choices of NBD as used in $(a)$.}
\label{F1a}
\end{figure}

In Fig. \ref{F1a} we show the results of attempts to fit both the experimentally measured (in the CMS experiment \cite{CMS}) multiplicity distributions, and the corresponding modified combinants $C_j$ calculated from these data. Note that these $C_j$ show very distinct oscillatory behavior (with a period roughly equal to $16$ in this case), which gradually disappears with $N$. It turns out that this oscillatory pattern cannot be reproduced by the $C_j$ calculated from a single NBD, we observe no trace of oscillations in this case. They begin gradually to appear for the $C_j$ calculated from $2$-NBD fits (with parameters from \cite{PG}) and become clearly visible when using $3$ component NBD (with parameters from \cite{Z}). In fact, in this case one can fit $C_j$ obtained from data \cite{Zborovsky}.

As shown in \cite{JPG,IJMPA} such oscillations of $C_j$ are seen for different pseudorapidity windows, in data from all LHC experiments and at all energies. The only condition is that the statistics of the experiment must be high enough, in cases of small statistics the oscillations become too fuzzy to be recognized \cite{IJMPA}. For the clarity of presentation we do not show errors on the figures with modified combinants $C_j$, leaving their discussion to Appendix \ref{Errors} (where we investigate the sensitivity of $C_j$'s to the uncertainties of the measurements).  Actually, a single NBD is not able to reproduce data because in this case the corresponding $C_j$ behave as
\begin{equation}
C_j = \frac{k}{\langle N\rangle} p^{j+1}, \label{CjNBD}
\end{equation}
i.e., all $C_j > 0$ \cite{JPG,IJMPA}. Quite contrary to the NBD, the modified combinants for the BD, cf. Eq. (\ref{BD}), oscillate rapidly,
\begin{equation}
C_j = (-1)^j \frac{K}{\langle N\rangle} \left( \frac{p}{1 - p}\right)^{j+1}, \label{C_jBD}
\end{equation}
with a period equal to $2$. However, their general shape lacks the distinctive fading down feature of the $C_j$ observed experimentally. This means that BD used alone cannot explain data.

The modified combinants $C_j$ defined by the recurrence relation (\ref{rCj}) are closely related to the {\it combinants} $C^{\star}_j$ introduced long a time ago in \cite{Combinants,Combinants-1} by means of the generating function, $G(z)=\sum^{\infty}_{N=0} P(N) z^N $, as
\begin{eqnarray}
C^{\star}_j &=& \frac{1}{j!} \frac{d^j \ln G(z)}{d z^j}\bigg|_{z=0} \label{C_j_star}\\
&{\rm or}&  \ln G(z) = \ln P(0) + \sum^{\infty}_{j=1} C^{\star}_j z^j \label{CombDef}
\end{eqnarray}
(see also \cite{Kittel,Book-BP,CombUse1,CombUse1a,CombUse1b,CombUse2,CombUse2a,CombUse2b,CombUse3,CombUse4,CombUse5}). Namely,
\begin{equation}
C_j = \frac{j+1}{\langle N\rangle} C^{\star}_{j+1}. \label{connection}
\end{equation}
Therefore, the $C_j$ can also be expressed by the generating function $G(z)$ of $P(N)$ as
\begin{equation}
\langle N\rangle C_j = \frac{1}{j!} \frac{ d^{j+1} \ln G(z)}{d z^{j+1}}\bigg|_{z=0}. \label{GF_Cj}
\end{equation}
This relation will be particularly useful later for calculation of the $C_j$ from the compound multiplicity distributions defined by some generating function $G(z)$. Note that, although the combinants, $C_j^{*}$, were already known for a long time, and their possible oscillatory behavior was also known, they have so far scarcely been used and were not directly extracted from the experimental data \cite{EVP,CombUse1,CombUse1a,CombUse1b,CombUse2,CombUse2a,CombUse2b,CombUse3,CombUse4,CombUse5}.

As in the case of the combinants, $C^{\star}_j$,  the set of modified combinants, $C_j$, provides a similar measure of fluctuations as the set of cumulant factorial moments, $K_q$, which are very sensitive to the details of the multiplicity distribution and were frequently used in phenomenological analyses of data (cf., \cite{Kittel,Book-BP}),
\begin{equation}
K_q = F_q - \sum_{i=1}^{q-1}\binom{q-1}{i-1} K_{q-i}F_i, \label{cumfactmom}
\end{equation}
where
\begin{equation}
F_q = \sum_{N=q}^{\infty} N(N-1)(N-2)\dots(N-q+1)P(N), \label{factmom}
\end{equation}
are the factorial moments. The $K_q$ can be expressed as an infinite series of the $C_j$,
\begin{equation}
K_q = \sum_{j=q}^{\infty}\frac{(j-1)!}{(j-q)!}\langle N\rangle C_{j-1}, \label{KconC}
\end{equation}
and, conversely, the $C_j$ can be expressed in terms of the $K_q$ \cite{Kittel,Book-BP},
\begin{equation}
C_j = \frac{1}{\langle N\rangle} \frac{1}{(j-1)!} \sum_{p=0}^{\infty}\frac{(-1)^p}{p!}K_{p+j}. \label{ConK}
\end{equation}
Modified combinants also share with cumulants the property of additivity. For a random variable composed of independent random variables, with its generating function given by the product of their generating functions, $G(x)=\prod_jG_j(x)$, the corresponding modified combinants are given by the sum of the independent components. On the other hand, while cumulants are best suited to the study of the densely populated region of phase space,  combinants are better suited for the study of sparsely populated regions because, according to Eq. (\ref{rCj}), calculation of $C_j$ requires only a finite number of probabilities $P(N<j)$ (wich may be advantageous in applications).

\begin{figure}[h]
\begin{center}
\includegraphics[scale=0.48]{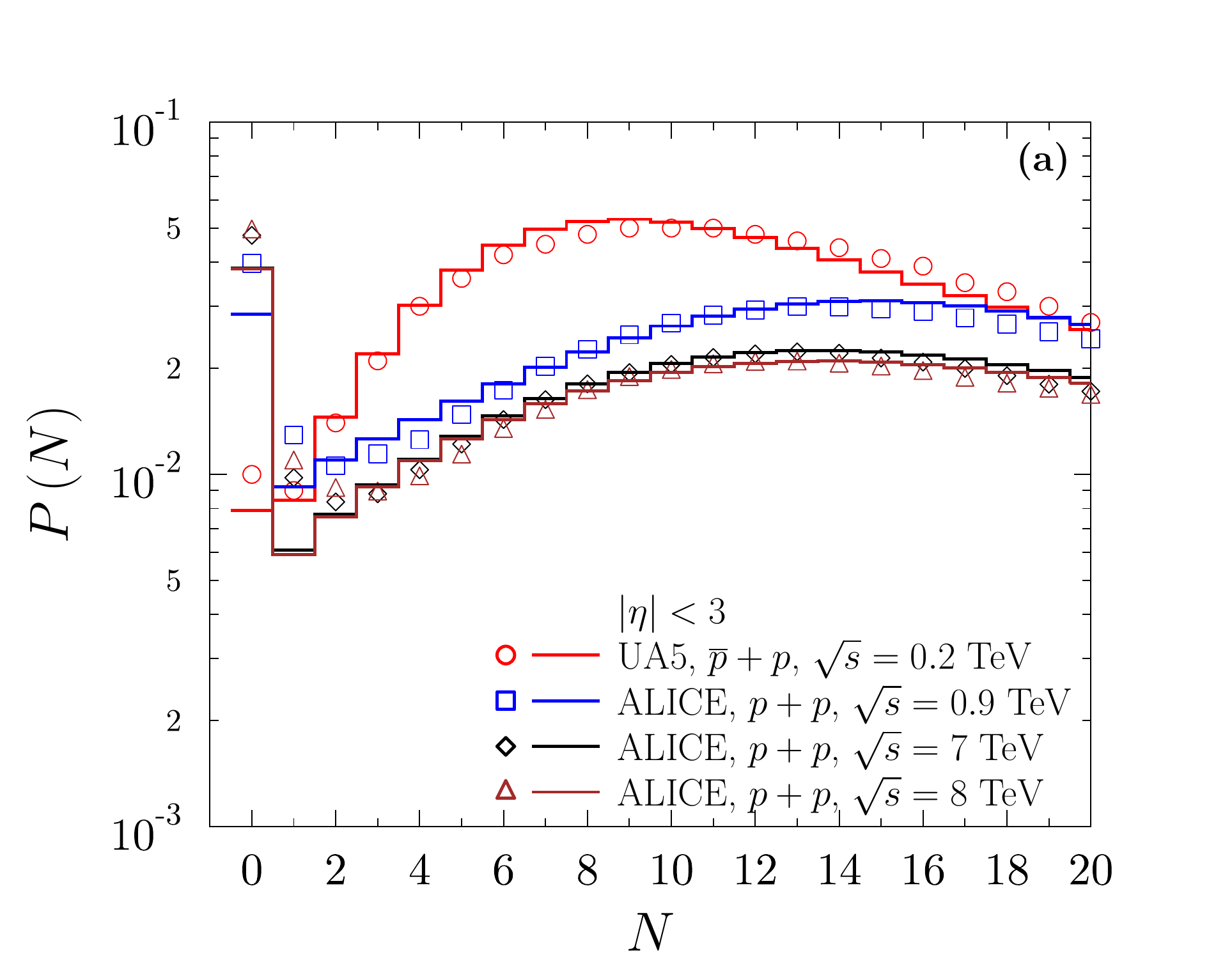}\\
\includegraphics[scale=0.48]{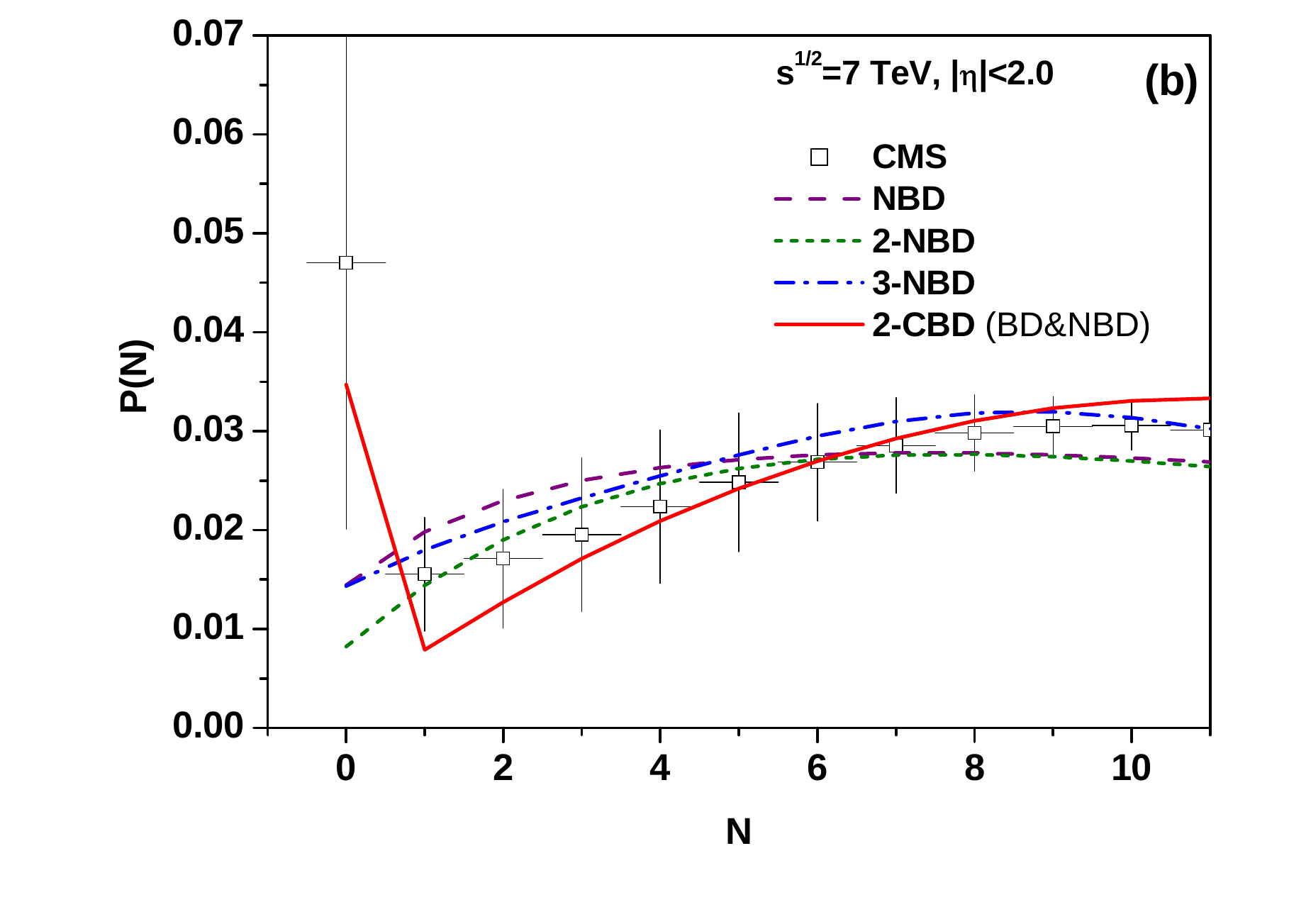}
\end{center}
\vspace{-7mm}
\caption{(Color online) $(a)$ Experimental smooth multiplicity distributions $P(N)$ displayed for low multiplicities and for energies ranging from $0.2$ TeV ($\bar{p}p$ collisions at UA5 experiment \cite{UA5}) up to $8$ TeV ($pp$ collisions at ALICE experiment \cite{ALICE}). Note the peculiar enhancement of the void probability $P(0)$ (rather small at $0.2$ TeV but quite substantial at $8$ TeV). $(b)$ The same as in $(a)$ but limited to small multiplicities to expose the void enhancement seen in $(a)$; it can be reproduced only by using a $2$-component compound binomial distributions (CBD, composed of NB and NBD) introduced and discussed below in Sections \ref{Compound} and \ref{Results}.}
\label{F1b}
\end{figure}

Concerning the void probabilities, the problem also to be addressed in this work is that, as can be seen in Fig. \ref{F1a} $(a)$, in the data on $P(N)$ discussed in this work one observes that $P(0) > P(1)$. As can be seen in Fig. \ref{F1b} $(a)$, such behavior occurs at all energies of interest. The interesting point is that it cannot be fitted either by a single NBD or by the compositions of $2$ or $3$ NBD used to fit the data presented in Figs. \ref{F1a}. However, as shown in Fig. \ref{F1b} $(b)$, this feature of the void probability can be nicely reproduced by a $2$-component compound distribution based on the BD and NBD, which we shall discuss in Sections \ref{Compound} and \ref{Results}.

Note that the void probability, $P(0)$, is strongly connected with the modified combinants. Using Eqs. (\ref{CombDef}) and (\ref{connection}) it can be written as:
\begin{equation}
P(0) = \exp\left( - \sum_{j=0}^{\infty} \frac{\langle N\rangle}{j+1}C_j\right). \label{P(0)}
\end{equation}
Using further Eq. (\ref{rCj}) one can show that the $P(0) > P(1)$ property is possible only when
\begin{equation}
\langle N\rangle C_0 < 1.\label{C0}
\end{equation}
For most multiplicity distributions we also have that $P(2) > P(1)$, which results in additional condition,
\begin{equation}
C_1 > C_0 \left( 2 - \langle N\rangle C_0\right), \label{C1C0}
\end{equation}
which together with Eq. (\ref{C0}) leads to the requirement that in this case
\begin{equation}
C_1 > C_0. \label{C1gtC0}
\end{equation}
However, this initial increase of $C_j$  cannot continue for all ranks $j$, rather, because of the normalization condition, $\sum_{j=0}^{\infty} C_j =1$, we should observe some kind of nonmonotonic behaviour of $C_j$ with rank $j$ in this case. Therefore, all multiplicity distributions for which the modified combinants $C_j$ decrease monotonically with rank $j$ (like, for example, the NBD, cf. Eq.(\ref{CjNBD}))  do not exhibit the enhanced void probability.

\section{Compound distributions}
\label{Compound}

Because a single distribution of the NBD or BD type cannot describe data we shall check the idea of {\it compound distributions} (CD). They are applicable when the production process consists of a number $M$ of some objects (clusters/fireballs/etc.) produced according to some distribution $f(M)$ (defined by a generating function $F(z)$), which subsequently decay independently into a number of secondaries, $n_{i = 1,\dots, M}$, following some other (always the same for all $M$) distribution, $g(n)$ (defined by a generating function $G(z)$). The resultant multiplicity distribution,
\begin{equation}
h\left( N =\sum_{i=0}^M n_i\right) = f(M)\otimes  g(n), \label{ftimesg}
\end{equation}
is a compound distribution of $f$ and $g$ with generating function
\begin{equation}
H(z) = F[G(z)]. \label{CD_GF}
\end{equation}
The immediate consequence of Eq. (\ref{CD_GF}) is that in the case where $f(M)$ is a Poisson distribution ($P_{PD}$ from Eq. ({\ref{PD})) with generating function
\begin{equation}
F(z) = \exp[\lambda (z - 1)], \label{GFPD}
\end{equation}
then, for any other distribution $g(n)$ with generating function $G(z)$, the combinants obtained from the compound distribution $h(N) = P_{PD} \otimes g(n)$ and calculated using Eq. (\ref{GF_Cj}), do not oscillate and are equal to
\begin{equation}
C_j = \frac{\lambda (j+1)}{\langle N\rangle}g(j+1). \label{Cjcalculated}
\end{equation}
In particular, in the case when $g(n)$ is a logarithmic distribution, $g(n) = - p^n/[n\ln(1-p)]$, for which $h(N)$ is the NBD with $k= - \lambda /\ln(1-p)$, one gets that the above $C_j$ coincide with those derived before from the recurrence relation (\ref{rCj}) and given by Eq. (\ref{CjNBD}). This reasoning can be further generalized to all more complicated compound distributions, with any distribution itself  being a compound poisson distribution. This limits the set of distributions $P(N)$ leading to oscillating $C_j$ only to, essentially, a BD and to all compound distributions based on it.

\begin{figure}[b]
\begin{center}
\includegraphics[scale=0.5]{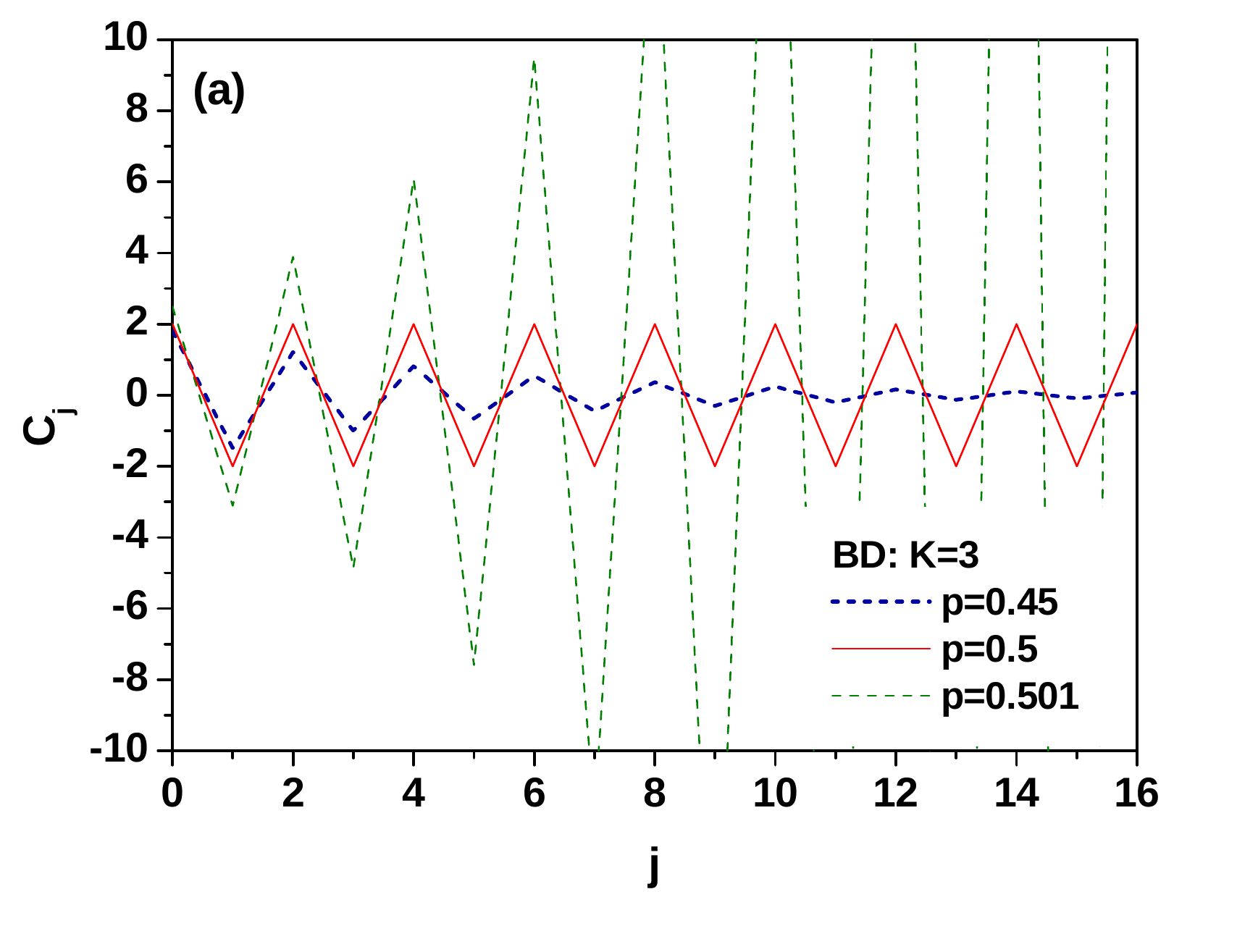}\\
\includegraphics[scale=0.5]{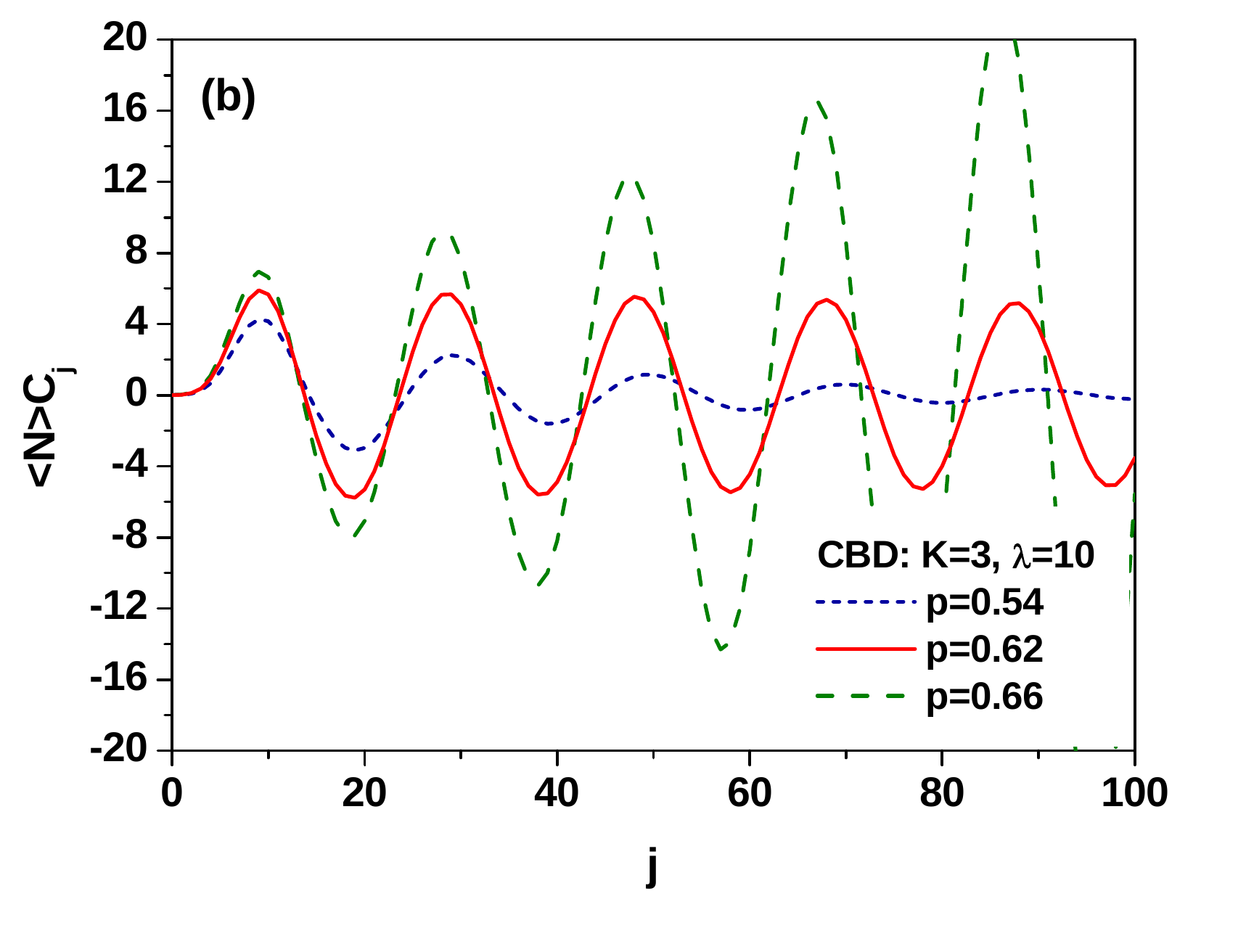}
\vspace{-2mm}
\end{center}
\vspace{-2mm}
\caption{(Color online) $(a)$ $C_j$ for a single BD for different probabilities of particle emission. $(b)$ The same BD compounded with a Poison distribution with $\lambda =10$.}
\label{F2a}
\end{figure}

It is interesting to note that this result explains the evident success of the multi-NBD type of $P(N)$ in fitting data on the $C_j$ \cite{Zborovsky}. Such distribution has freely selected weights and parameters $(p,k)$ of NBDs and apparently looks similar to the compound distribution of the BD with the NBD discussed in the next Section, $(BD\&NBD)$, which is also a multi-NBD distribution but this time its weights are precisely given by the BD, and parameters $(p,k)$ of each NBD component are also fixed. Note that the sum of $M$ variables (with $M = 0,~1,~2,\dots$), each from the NBD characterized by parameters $(p,k)$, is described by a NBD characterized by $(p,Mk)$. In the case where $M = 0,~1,\dots, K$ is distributed according to a BD, we have a $K$-component NBD (where consecutive NBD have precisely defined parameters $k$),
\begin{equation}
P(N) = \sum_{M=0}^K P_{BD}(M) P_{NBD}(N;p,Mk), \label{BD-NBD}
\end{equation}
which naturally leads to the appearance of oscillations. Note that in this case one has also $M=0$ component, which is lacking in the usual multi-NBD approach. This is the reason why the compound $(BD\&NBD)$ distribution reproduces the void probability, $P(0)$, while the single NBD, or any combination of NBDs, do not. As for the modified combinants it only changes their amplitude and periods of oscillations.

\begin{figure}[b]
\begin{center}
\includegraphics[scale=0.5]{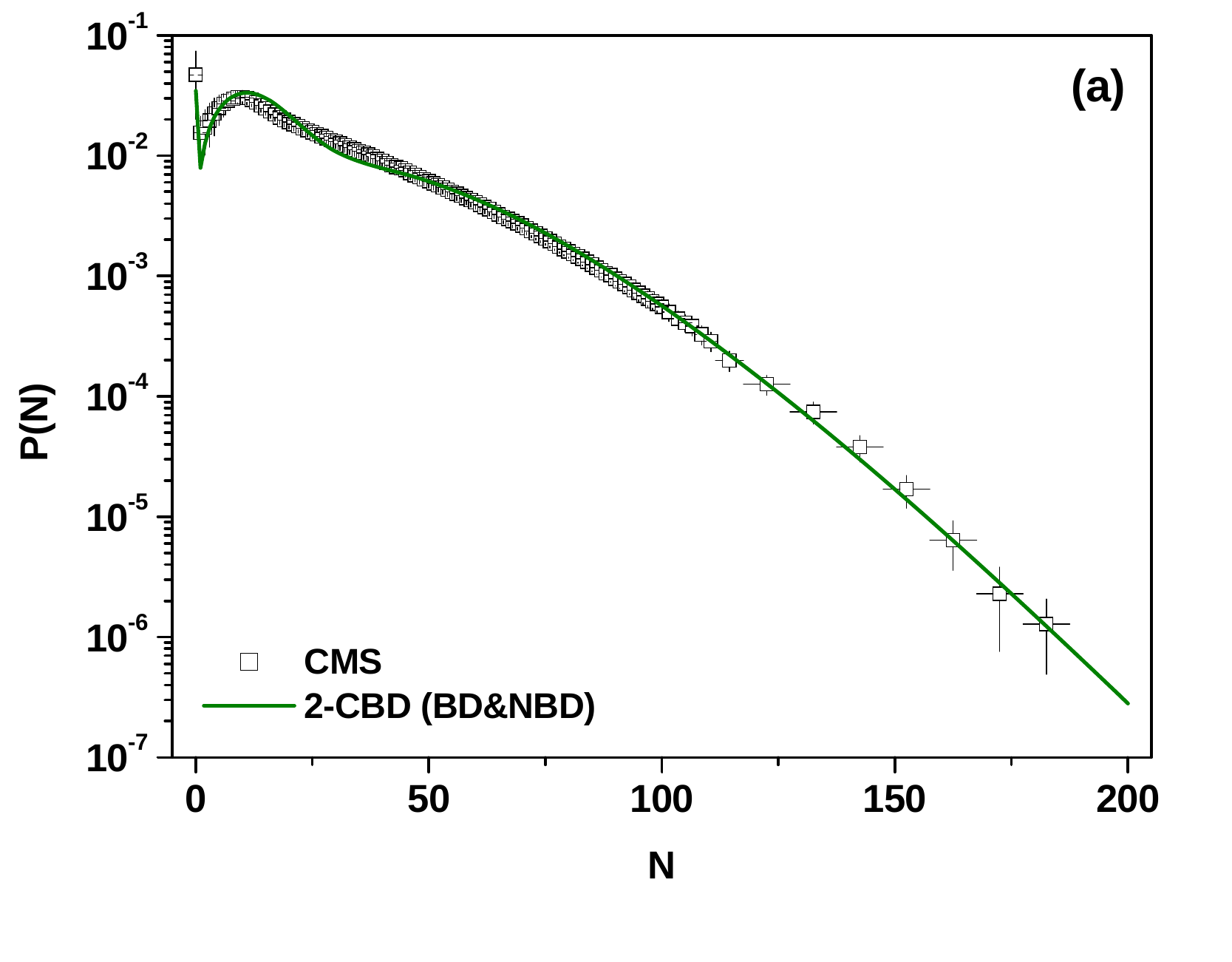}\\
\includegraphics[scale=0.5]{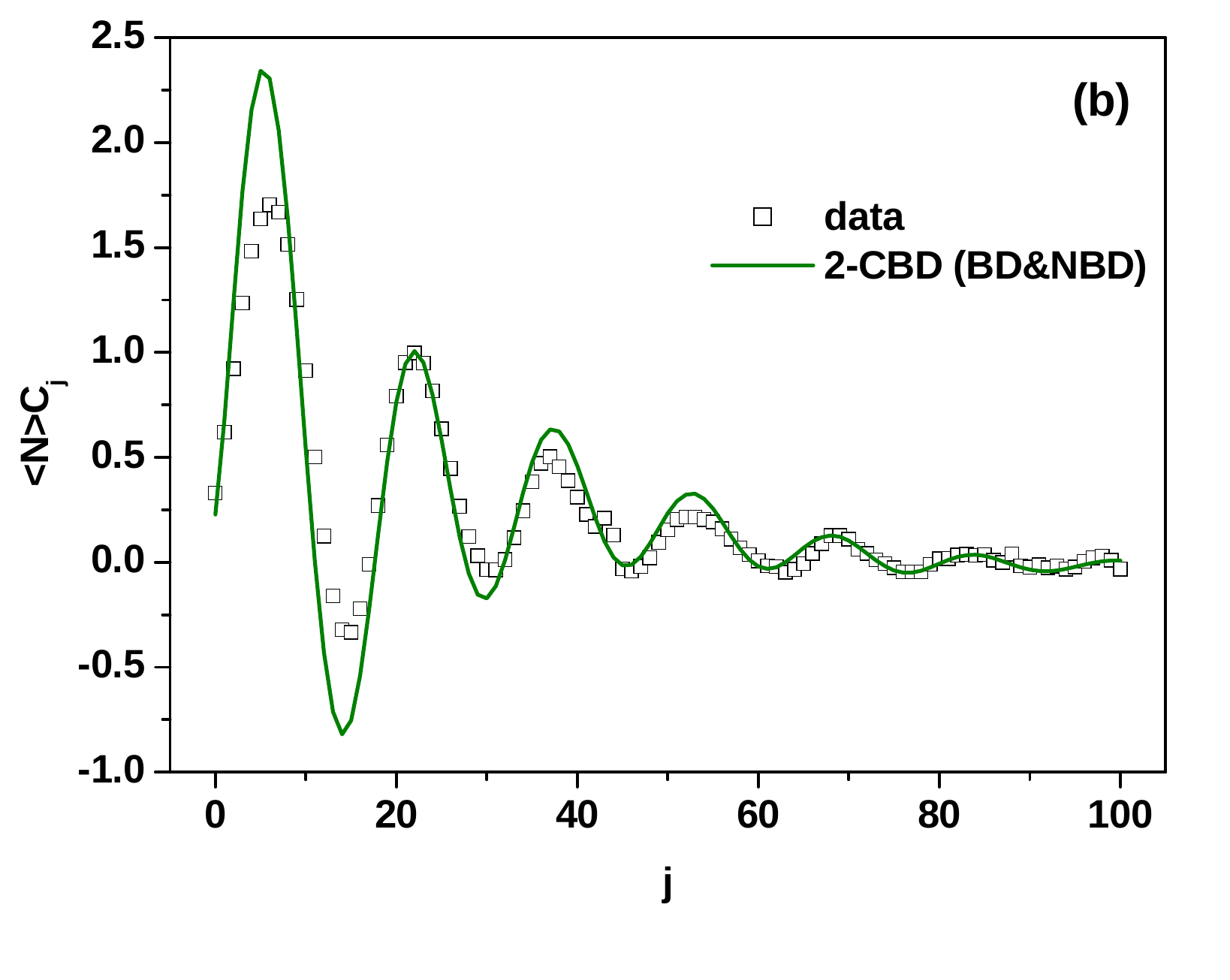}
\end{center}
\vspace{-2mm}
\caption{(Color online) $(a)$  Fits to CMS data \cite{CMS} for $P(N)$. $(b)$ Fits to the $C_j$ obtained from the same data using a combination of two CBD.}
\label{F2b}
\end{figure}

\section{Results}
\label{Results}

As mentioned above, the modified combinants $C_j$ for the BD with generating function
\begin{equation}
F(z) = (pz + 1 - p)^K \label{GFBD}
\end{equation}
oscillate with a period of $2$, whereas, as shown in Fig. \ref{F2a} $(a)$, the amplitudes of these oscillations depend on the probability emission $p$. To control the period of the oscillations one has to compound this BD with some other distribution. Fig. \ref{F2a} $(b)$ shows an example of using for this purpose a Poisson distribution with a generating function given by Eq. (\ref{GFPD}) (for which $C_0 = 2$ and $C_{j>0} = 0$). The generating function of the resulting Compound Binomial Distribution (CBD) is
\begin{equation}
H(z) = \left\{ p \exp[ \lambda (z-1)] + 1 -p \right\}^K. \label{H_FG}
\end{equation}

\begin{figure}[b]
\begin{center}
\includegraphics[scale=0.49]{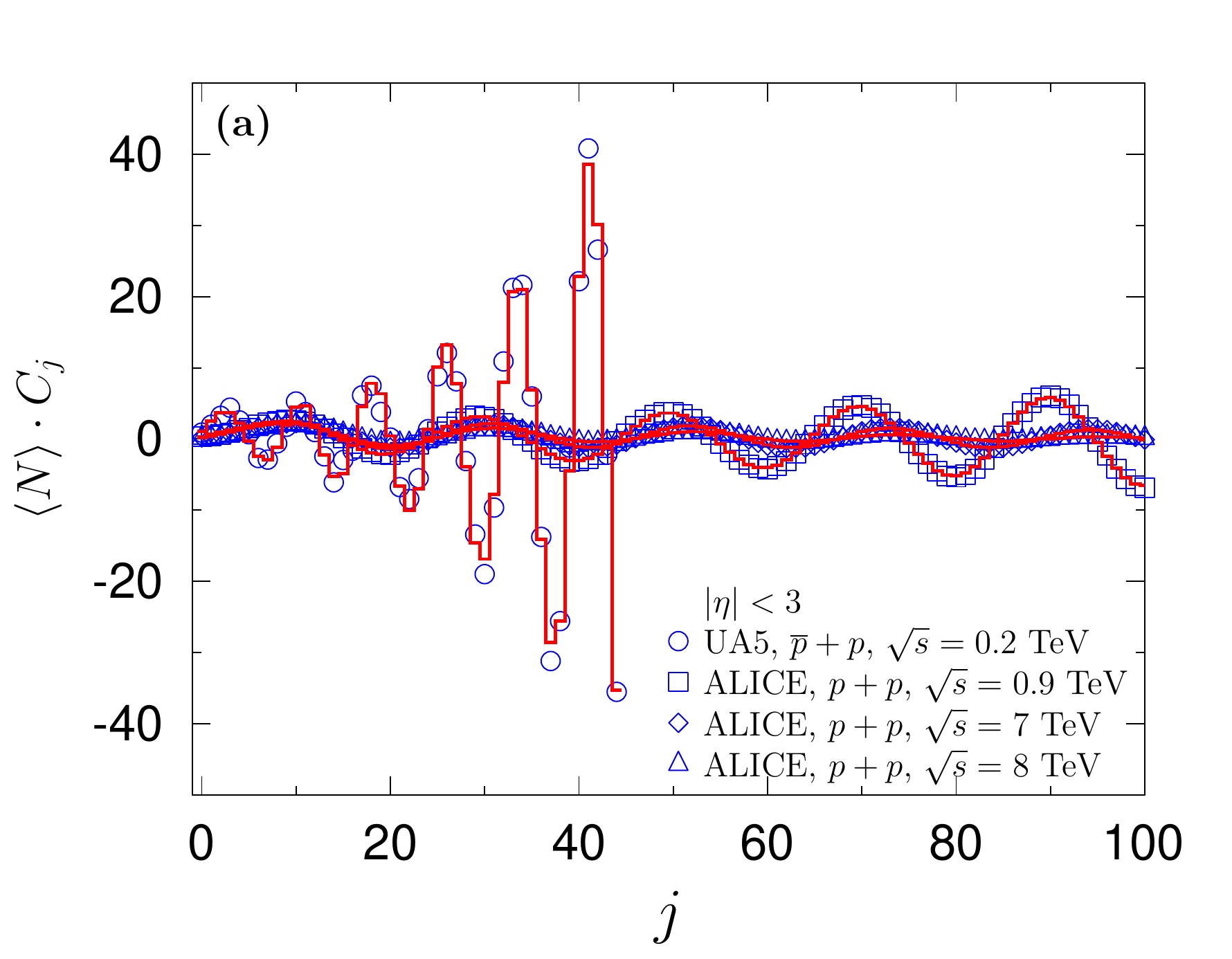}\\
\includegraphics[scale=0.49]{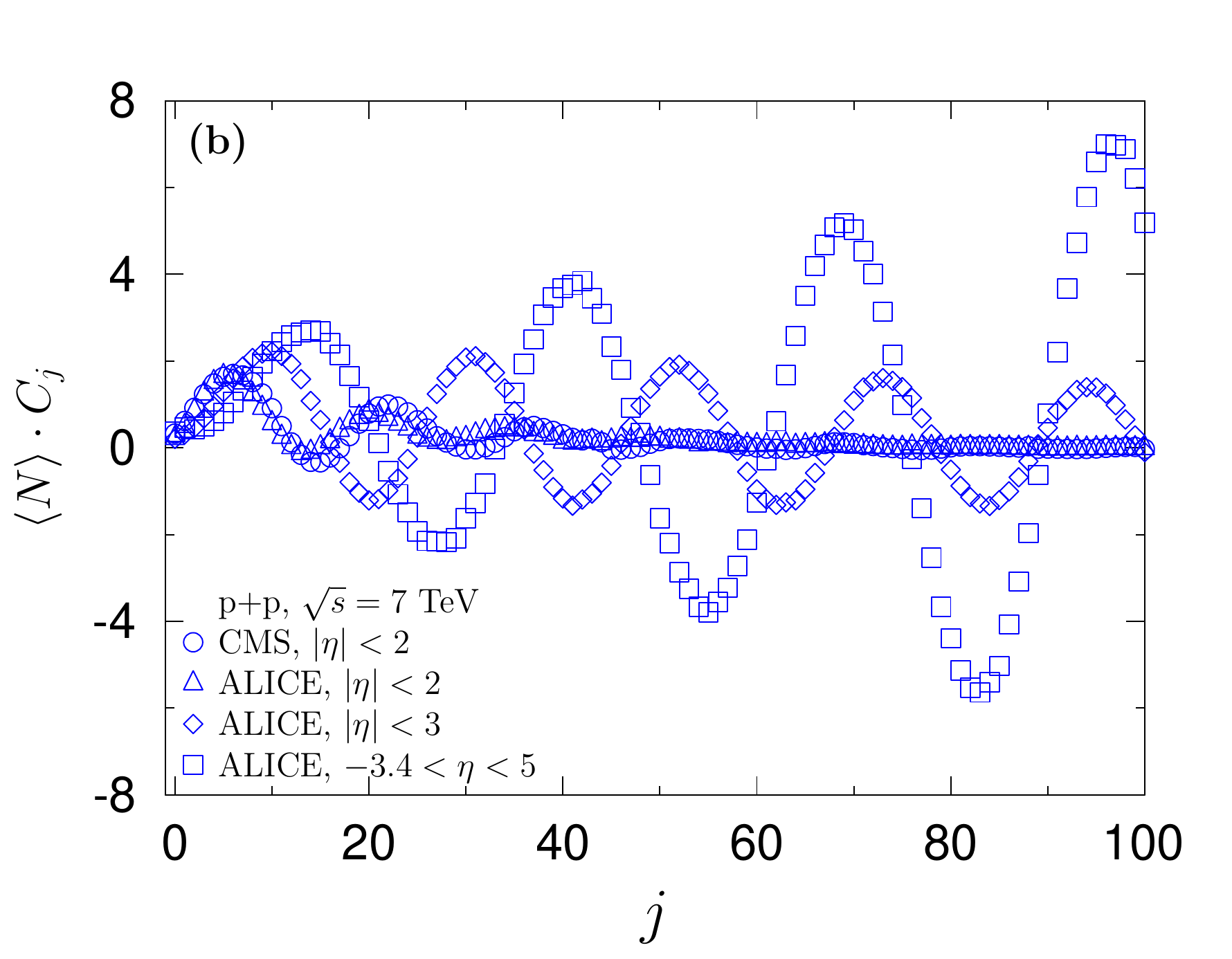}
\end{center}
\vspace{-2mm}
\caption{(Color online) $(a)$ $\langle N\rangle C_j$ for different energies (UA5 results are from \cite{UA5}). $(b)$ The same for different rapidity windows as measured by the CMS \cite{CMS} and ALICE \cite{ALICE} experiments.}
\label{F2c}
\end{figure}


\begin{figure*}[t]
\begin{center}
\includegraphics[scale=0.61]{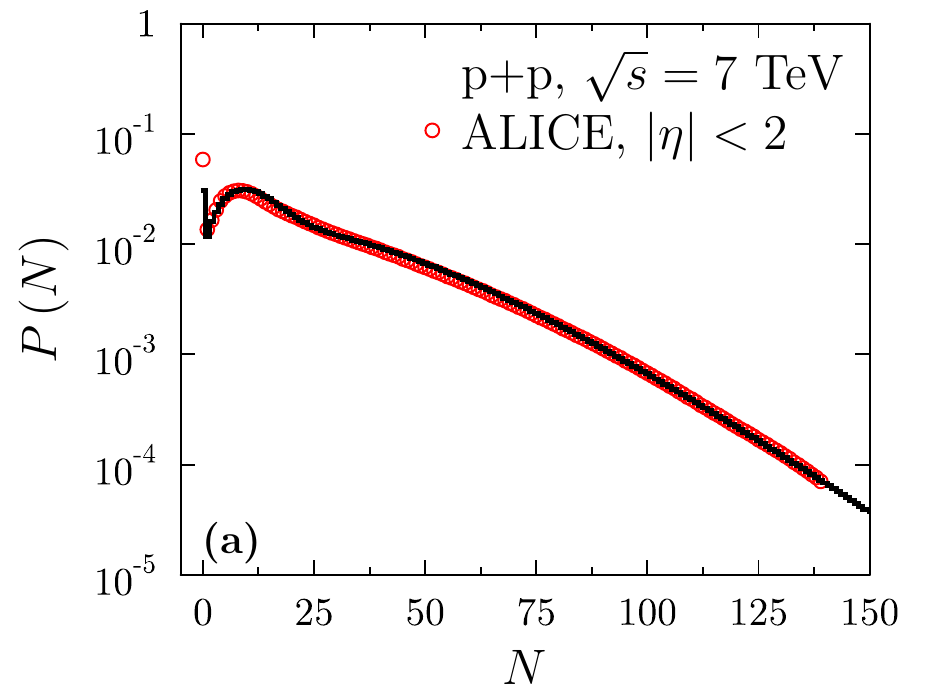}
\includegraphics[scale=0.61]{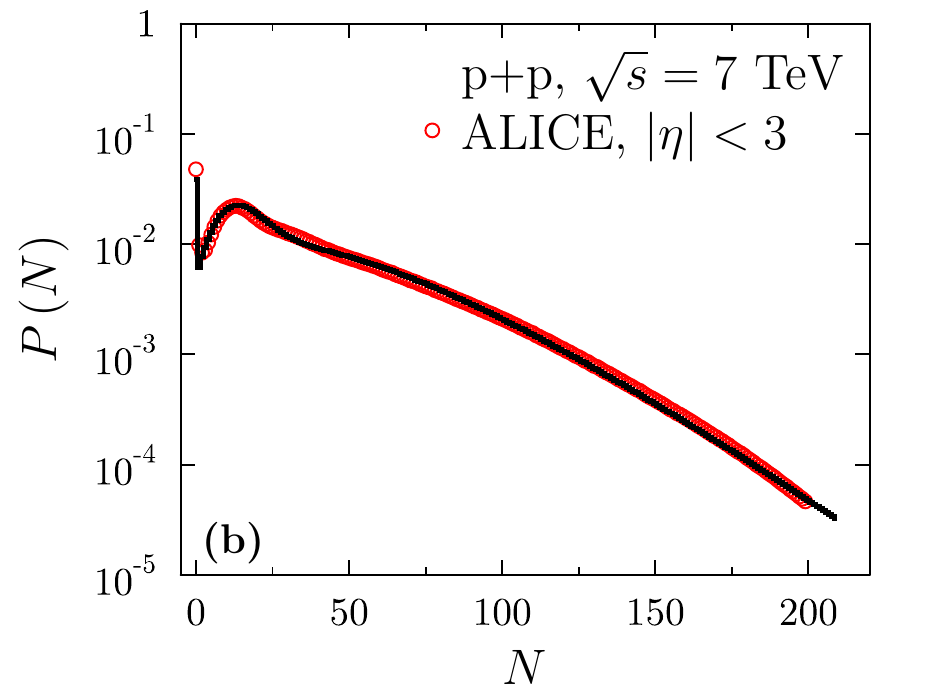}
\includegraphics[scale=0.61]{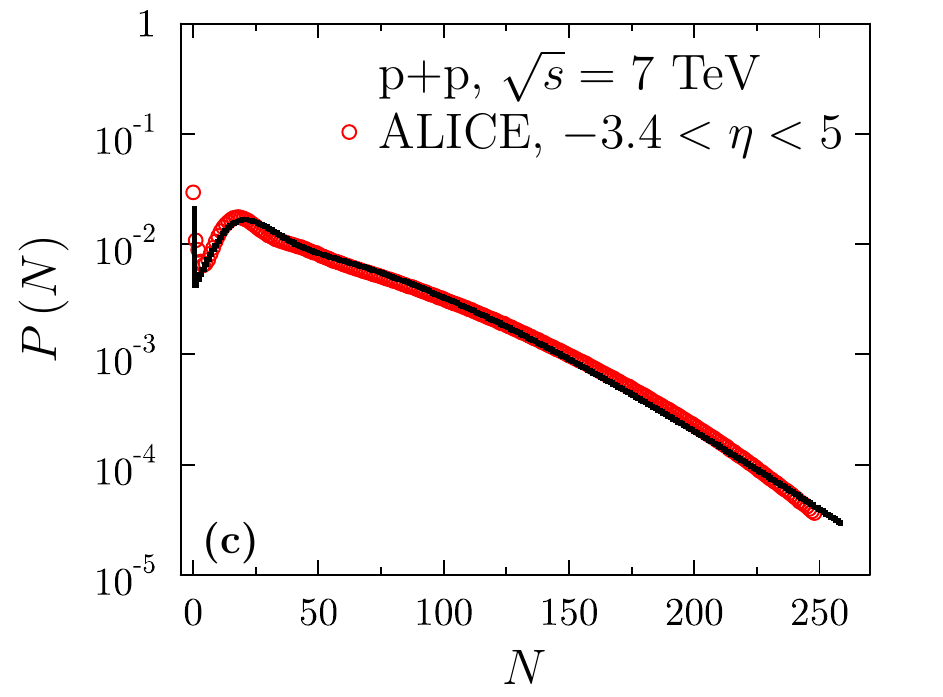}\\
\vspace{2mm}
\includegraphics[scale=0.61]{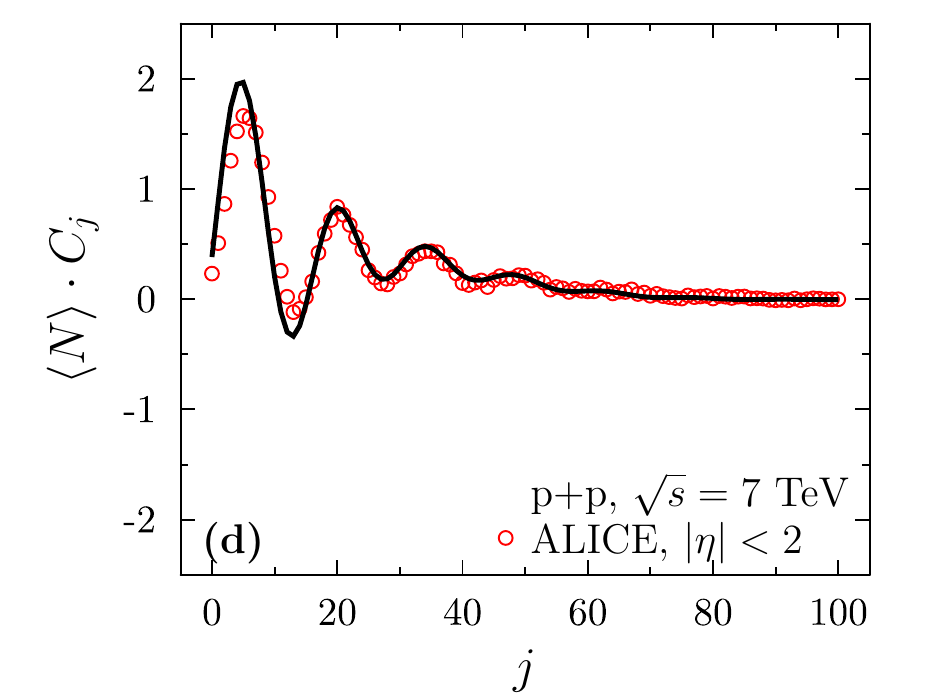}
\includegraphics[scale=0.61]{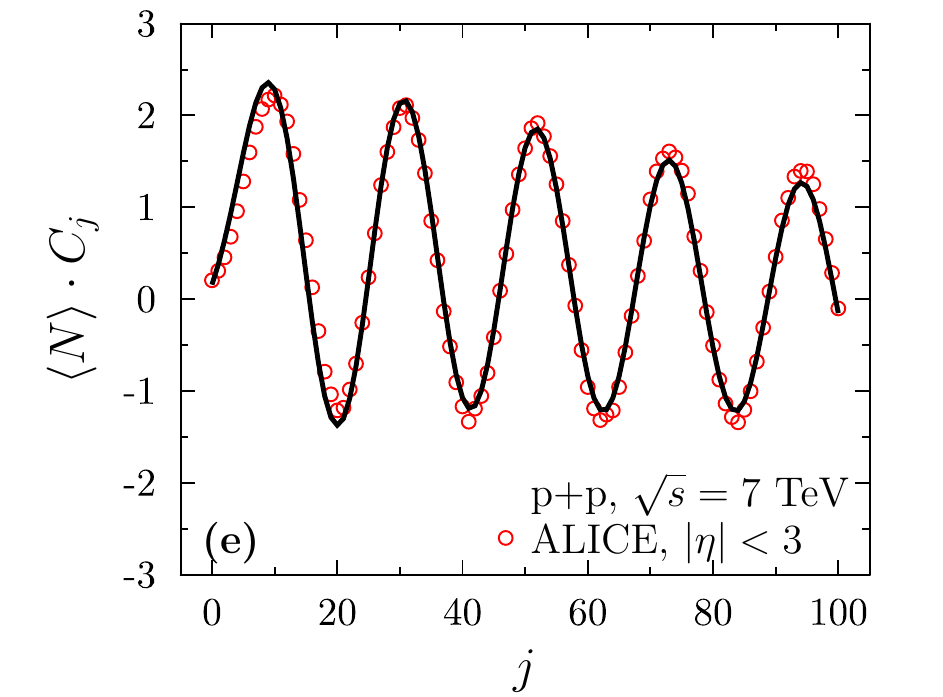}
\includegraphics[scale=0.61]{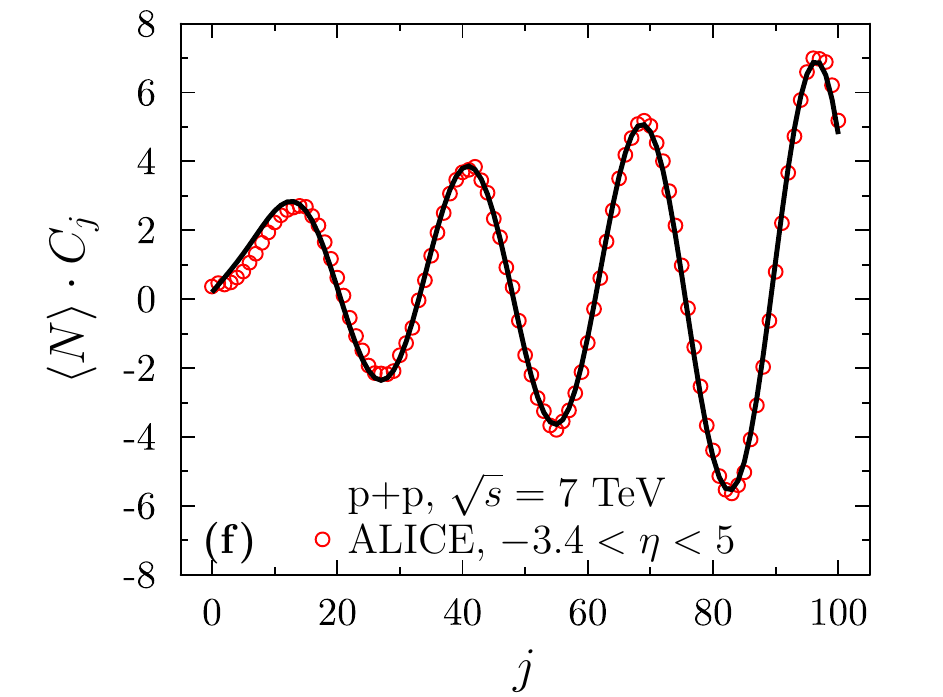}
\end{center}
\caption{(Color online) $(a)$ - $(c)$ Multiplicity distributions $P(N)$ measured by ALICE \cite{ALICE}. $(d)$-$(f)$ The corresponding modified combinants $C_j$ emerging from them fitted using a two compound distribution (BD+NBD) given by Eqs. (\ref{2-CBD}) and (\ref{2CBD}) with parameters listed in Table \ref{tab-1}.}
\label{F3}
\end{figure*}
\begin{table*}
\centering
\caption{Parameters $w_i$, $p_i$, $K_i$, $k_i$ and $m_i$ of the $2$-component $P(N)$, Eqs. (\ref{2-CBD}) and (\ref{2CBD}), used to fit the data in Fig. \ref{F3}. For completeness the corresponding $p'_i = m_i/(m_i + k_i)$ from Eq. (\ref{2-CBD}) are also included.}
\label{tab-1}
\vspace{2mm}
\begin{tabular}{|c|cccccc|cccccc|}
\hline
& & & & & & & & & & & & \\
                                   & $w_1$      & $p_1$  & $K_1$  & $k_1$ & $m_1$  & $p'_1$ & $w_2$  & $p_2$  &$K_2$& $k_2$ & $m_2$ & $p'_2$~  \\
                                   & & & & & & & & & & & & \\ \hline
                                   & & & & & & & & & & & & \\
 $-2\langle~\eta~\langle 2$~~       &~ $0.30$    & $0.75$  & $3$    & $3.8$ & $4.75$ & $0.56$~ &~ $0.70$ & $0.70$ & $3$ & $1.30$ & $15.9$ & $0.924$~ \\
 & & & & & & & & & & & &  \\
 $-3\langle~\eta~\langle 3$~~        &~ $0.24$    & $0.90$  & $3$    & $2.8$ & $5.75$ & $0.67$~ &~ $0.76$ & $0.645$ & $3$ & $1.34$ & $23.5$ & $0.946$~ \\
 & & & & & & & & & & & & \\
 $-3.4\langle~\eta~\langle 5$~      &~ $0.20$    & $0.965$  & $3$    & $2.7$ & $8.00$ & $0.75$~ &~ $0.80$ & $0.72$ & $3$ & $1.18$ &
 $27.0$ & $0.955$~ \\
 & & & & & & & & & & & & \\ \hline
\end{tabular}
\end{table*}

Fig. \ref{F2a} $(b)$ shows the $C_j$ obtained from such a CBD with $K=3$ and $\lambda =10$ and calculated for three different values of $p$ in the BD: $p=0.54,~0.62,~0.66$. Note that, in general, the period of oscillation is now equal to $2\lambda$ (i.e., here, where $\lambda = 10$, it is equal to $20$). However, such a CBD lacks the fading down feature of its $C_j$ and therefore cannot fit the results presented here. The situation improves substantially when one uses a multi-CBD based on Eq. (\ref{H_FG}), but still the agreement is not satisfactory. The situation improves dramatically if one replaces the Poisson distribution by a NBD and, additionally, allows a two-component version of such a CBD in order to gain better control over both the period and the amplitudes and on their behavior as a function of the rank $j$, and
\begin{equation}
P(N) = \sum_{i=1,2} w_i h\left(N; p_i, K_i, k_i, m_i\right);\quad \sum_{1=1,2} w_i = 1.  \label{2CBD}
\end{equation}
The generating function of such a CBD is
\begin{equation}
H(z) = \left[ p\left( \frac{1 - p'}{1 - p'z}\right)^k + 1 - p\right]^K,~~~~p' = \frac{m}{m + k}. \label{2-CBD}
\end{equation}
As one can see in Fig. \ref{F2b} in this case (with  $K_1 = K_2 = 3$, $p_1 = 0.7$, $p_2 = 0.67$, $k_1 = 4$, $k_2 = 2.3$, $m_1 = 6$, $m_2 = 19.0$ and $w_1 = w_2 = 0.5$) one can nicely fit both the $P(N)$ and $C_j$.  Of special importance is the fact that the enhancement $P(0)>P(1)$ is also reproduced in this approach. This is best visible in Fig. \ref{F1b}$(b)$ which concentrates on the region of small $N$ only. This result is presented there in comparison with the results of a number of other, seemingly very good, fits based only on some combinations of NBD (and not using the BD), which are not able to reproduce this feature of the data.

Summarizing this part: it turns out that to describe all aspects of the data on multiparticle distributions one has to use a multicomponent compound distribution based on the BD (which is responsible for the oscillations in $C_j$) which is compounded with some other distribution providing damping of the oscillations for large $N$ (in this example it is a NBD). In Fig. \ref{F2c} $(a)$ we show the results of such approach applied to data taken at different energies (in rapidity window $|\eta| < 3$).

Fig. \ref{F2c} $(b)$ shows another intriguing property of the modified combinants, namely that both their periods and amplitudes increase with the width of the rapidity window in which the data on the resulting $P(N)$ were collected. A more detailed picture of this phenomenon is presented in Fig. \ref{F3} showing the description of the multiplicity distributions (NSD events at $7$ TeV) measured by ALICE \cite{ALICE} for three different rapidity windows: $|\eta| < 2$, $|\eta|<3$ and $-3.4<\eta<5$. The most intriguing feature observed is the rather dramatic increase of both the period of the oscillations and their amplitude with the width of the rapidity window used to collect the data and, most noticeably, the previously observed fading down of their amplitude is now replaced by an (almost) constant behavior (for $|\eta| < 3$) or by a rather dramatic increase (for $-3.4 < \eta < 5$). Because $\langle N\rangle \sim \Delta \eta $, some part of this increase could be caused by the increase of $\langle N\rangle$ with $\Delta \eta$, the rest expects an explanation. However, in general, with increasing $\Delta \eta$ both probabilities, $p$ and $p'$, are increasing which results in an increase in the amplitudes of the $C_j$ with rank $j$.

So far both components are based on the same BD with $K_{1,2}=3$. We have checked that one can safely increase the parameter $K_2$ in the second component while keeping practically all the parameters of the first component the same and, for appropriately selected other parameters of the second component, we get results essentially indistinguishable from those presented in Fig. \ref{F3} (for example, for the ALICE data at $7$ TeV and $|\eta|<2$, this can be done for $K_2 = 3,~6,~9,~~12,20$ and $\left( p_2,~m_2\right)$ such that $K_2 p_2 m_2 = const \simeq 32$, which means that for the second component $\langle N_{BD}\rangle\langle N_{NBD}\rangle \simeq 32$). So far we cannot offer any convincing explanation of our findings. At the moment the rough idea could be, for example, that the two components correspond to a quark-quark interaction (therefore $K = 3$) and to a gluon-gluon interaction (in this case $K$ could be different, as mentioned above)\footnote{To be more specific, one can try to estimate the number of a "hard" gluons participating in the interaction (which is equivalent to $K_2$). For example, in \cite{ZW} it was estimated as $N_G = 2.84 \ln(\sqrt{s}) - 11.45$, which for an energy of $7$ TeV gives $N\sim 14$.}.

\section{Summary and conclusions}
\label{sum-concl}

\begin{figure*}
\begin{center}
\includegraphics[scale=0.61]{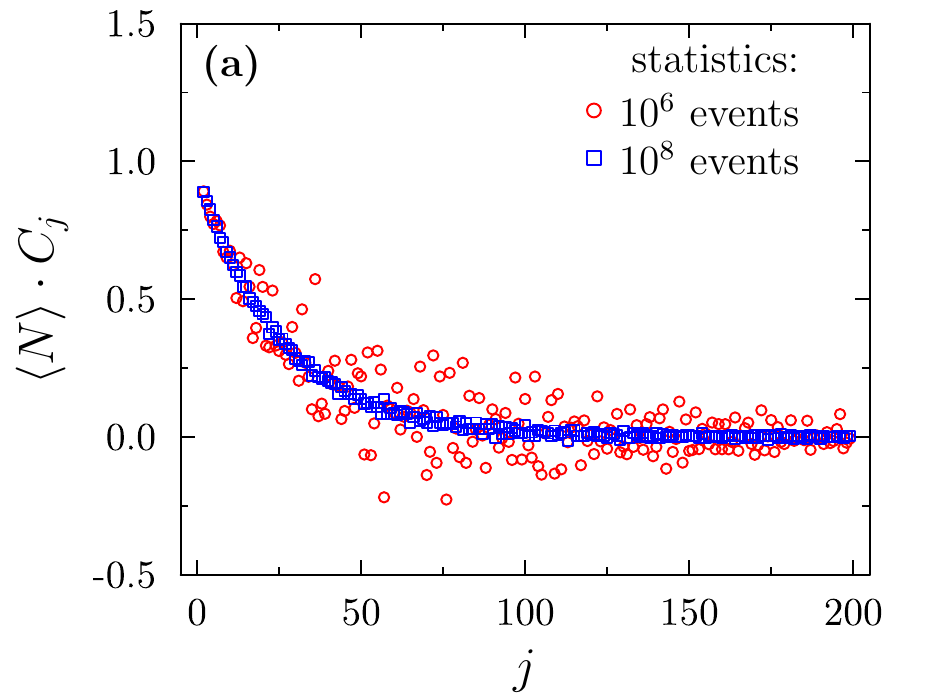}
\includegraphics[scale=0.61]{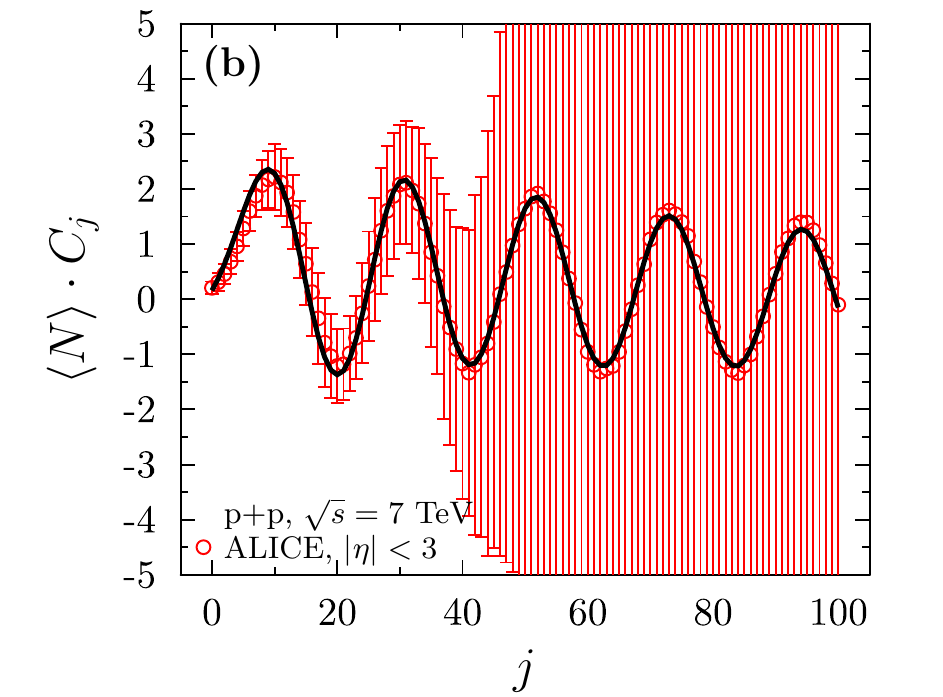}
\includegraphics[scale=0.61]{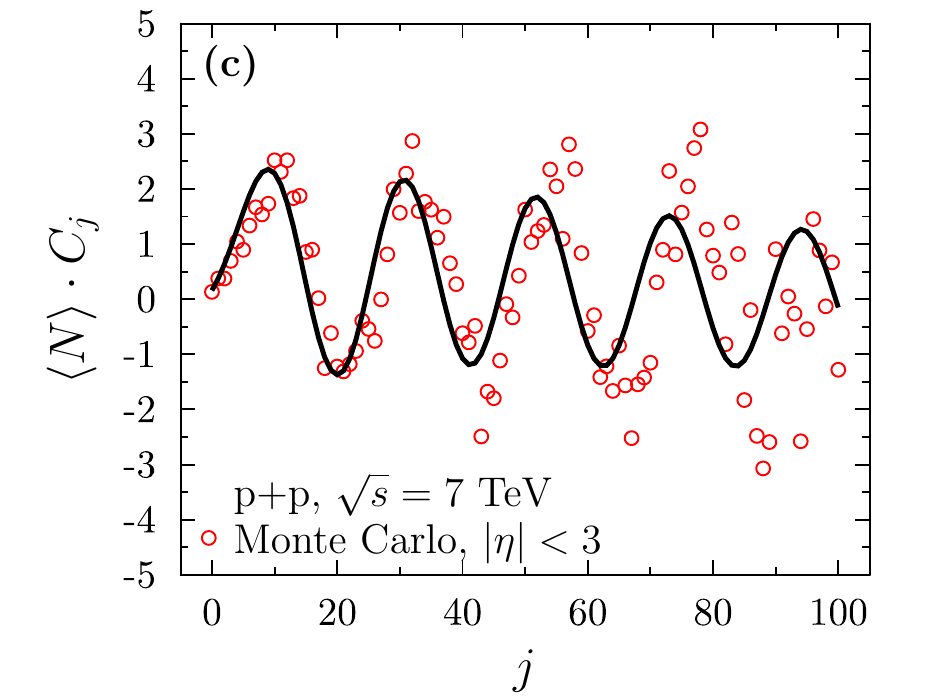}\\
\vspace{2mm}
\includegraphics[scale=0.61]{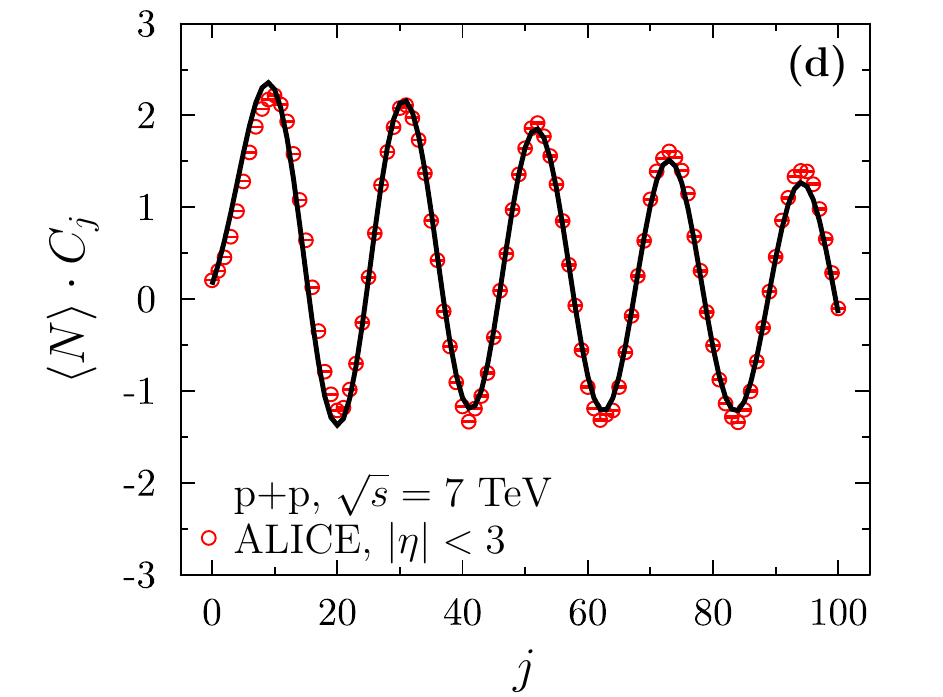}
\includegraphics[scale=0.61]{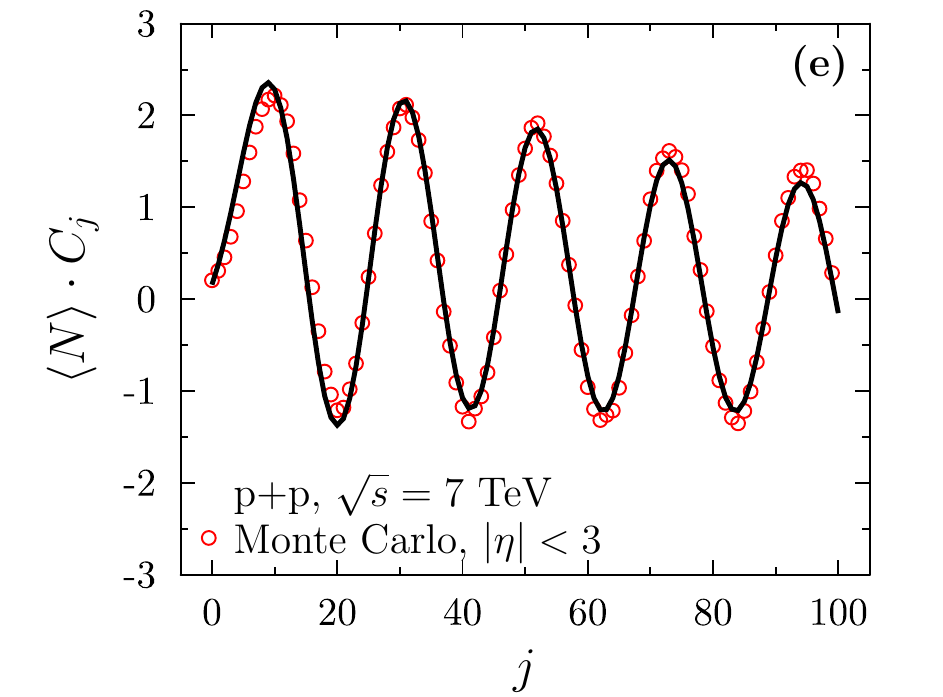}
\includegraphics[scale=0.61]{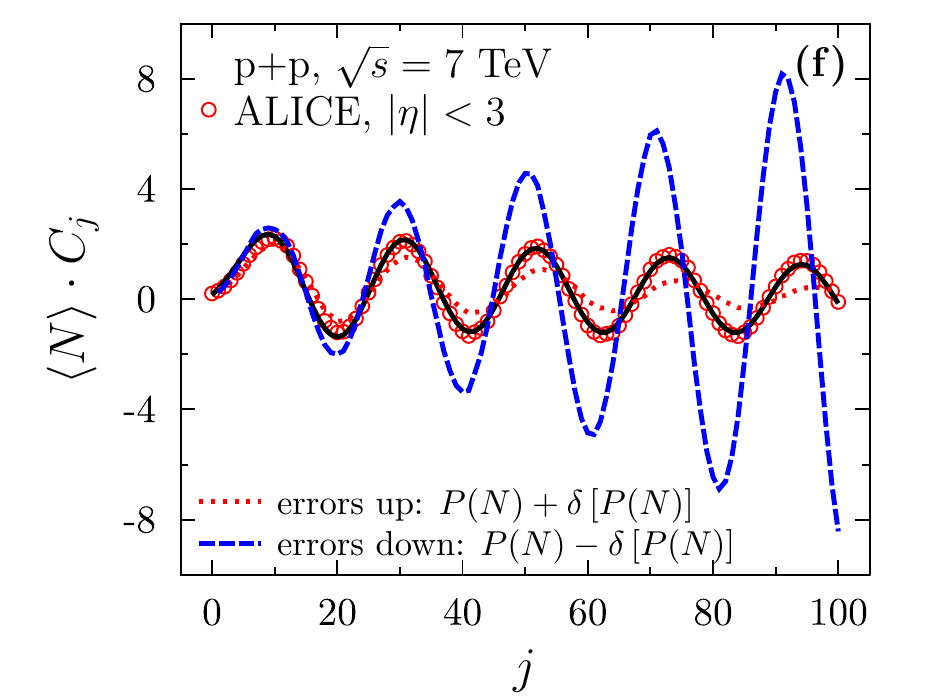}
\end{center}
\vspace{-5mm}
\caption{(Color online) $(a)$ Monte Carlo evaluated coefficients $C_j$ emerging from NBD with parameters: $\langle N\rangle = 25.5$ and $k = 1.45$. With increasing statistics points are merging to a continuous line.
$(b)$ Errors of $\langle N\rangle C_j$ evaluated using the systematic and statistical uncertainties of $P(N)$ given by ALICE \cite{ALICE}.
$(c)$ Monte Carlo evaluated coefficients $\langle N\rangle C_j$ emerging from the systematic and statistical errors of $P(N)$. The curve presented here denotes the fit to the original coefficients $C_j$ obtained from the measured $P(N)$, it is not the fit to the points shown.
$(d)$ For the same data as before the errors were evaluated assuming only statistical uncertainties of the measured $P(N)$ with a poissonian distribution of events in each bin, i.e., $Var[P(N)] = P(N)/N_{stat}$. Note that in this case statistical errors do not give any noticeable errors of $C_j$.
$(e)$ Monte Carlo evaluated coefficients $\langle N\rangle C_j$ with only statistical errors of $P(N)$ accounted for.
The continuous curve represents the fit to the original coefficients $C_j$ obtained from the measured $P(N)$.
$(f)$ The modified combinants $C_j$ emerging from the ALICE data on $P(N)$ \cite{ALICE} (continuous curve) in envelope corresponding to the systematic uncertainties of data, $P(N) \pm \delta[P(N)]$.}
\label{Fig-errors}
\end{figure*}

We presented an approach in which one can simultaneously reproduce such features of the observed  multiplicity distributions as: their shape as a function of the multiplicity, $P(N)$, the peculiar properties of the observed void probabilities, $P(0) > P(1)$, and, finally, the behavior of the modified combinants, $C_j$, which can be deduced from the measured $P(N)$. In particular, we have shown that the most popular type of multiplicity distribution, the NBD,  cannot alone describe the data on $P(N)$. The $2$-component NBD can describe the data on $P(N)$ but fails to describe the observed oscillatory pattern of the $C_j$ obtained from them. These two features can be fully reproduced by multicomponent NBD models  (like, for example, the $3$-NBD proposed in \cite{Zborovsky}), but such models do not reproduce the property that $P(0) > P(1)$. In fact, none of this class of models reproduces it (as long as it does not take into account also contributions from the single-diffractive and double-diffractive events \cite{Adam-1}). On the other hand, the compound distributions discussed in \cite{EVP,Void}, which were specially designed and tuned to describe the $P(0) > P(1)$ property, do not reproduce the oscillatory behavior of the corresponding modified combinants, $C_j$. This is because they belong to the group of infinitely divisible distributions (for which $C_j$ are positive for all ranks $j$~\cite{Book-BP}).

Our approach is based on the compound distribution, CBD, with the main role played by the BD. It is responsible for the oscillatory behavior of the modified combinants, $C_j$, and is compounded with a NBD which is responsible for the amplitudes and periods of these oscillations. In the framework of  clusters description such compound distribution corresponds to the use of the binomial distribution for the cluster distribution and the negative binomial distribution as the distribution for the fragmentation of the clusters (the mean multiplicity of NBD determine the period of oscillations).

The lack of oscillatory behavior of the $C_j$ deduced from the NBD can then be attributed to the fact that the NBD is itself a compound distribution of the Poisson and logarithmic distributions, and compound distribution of a Poisson with any other distribution always results in non-oscillating (in fact, exponentially fading down) $C_j$. On the other hand, the emergence of the oscillatory behavior of the multi-NBD can be attributed to the fact that a sum of NBDs is, under some conditions, equivalent to a compound distribution of a BD with a NBD.

Of course, we are aware that many other models are used  to describe multiplicity distributions, but 40 years have already passed (counting from the Ref.~\cite{Combinants-1}) with no detailed experimental study of the combinants and with a rather sporadic attempts of their phenomenological use to describe the multiparticle production processes (cf., references~\cite{Kittel,JPG,IJMPA,Zborovsky,Book-BP,CombUse1,CombUse1a,CombUse1b,CombUse2,CombUse2a,CombUse2b,CombUse3,CombUse4,CombUse5,Hegyi}). Among them scenarios with the fluctuation of $\langle N\rangle$ in the Poisson distribution (formally correspond to the so called Poisson transforms) seem to be very promising. It is remarkable that fluctuation of $\langle N\rangle$ in the Poisson distribution, with $f\left(\langle N\rangle\right)$ given by the generalized gamma distribution, leads to fractional negative binomial distribution (known also as HNBD, because such extension of NBD can be expressed in terms of the Fix’s H-function) which demonstrates oscillatory behavior of the corresponding combinants~\cite{Hegyi,Hegyi:1997ty,Hegyi:1997im,Hegyi:1996bt}. Despite that in the HNBD we have $P\left(0\right)<P\left(1\right)$, such extension of NBD (with only one additional parameter) is worth future detailed study.

In summary, we believe that the modified combinants, $C_j$, deduced from the measured multiplicity distributions, $P(N)$, together with the already measured void probabilities, could provide additional information on the dynamics of the particle production. This, in turn, could allow us to reduce the number of possible interpretations presented so far and, perhaps, answer some of the many still open fundamental questions. Experimental measurements of $C_j$ (or, rather, presenting them together with the already measured $P(N)$), appear in this context as a new important necessity.

\vspace*{0.3cm}
\centerline{\bf Acknowledgements}
\vspace*{0.3cm}
This research  was supported in part by the Polish Ministry of Science and Higher Education (DIR/WK/2016/2010/17-1) and National Science Centre (DEC-2016/22/M/ST/00176) (GW) and by the National Science Center (NCN) under contract 2016/23/B/ST2/00692 (MR). We would like to thank Dr Nicholas Keeley for reading the manuscript.

\appendix*

\section{Estimation of errors}
\label{Errors}

\begin{figure}[t]
\begin{center}
\includegraphics[scale=0.8]{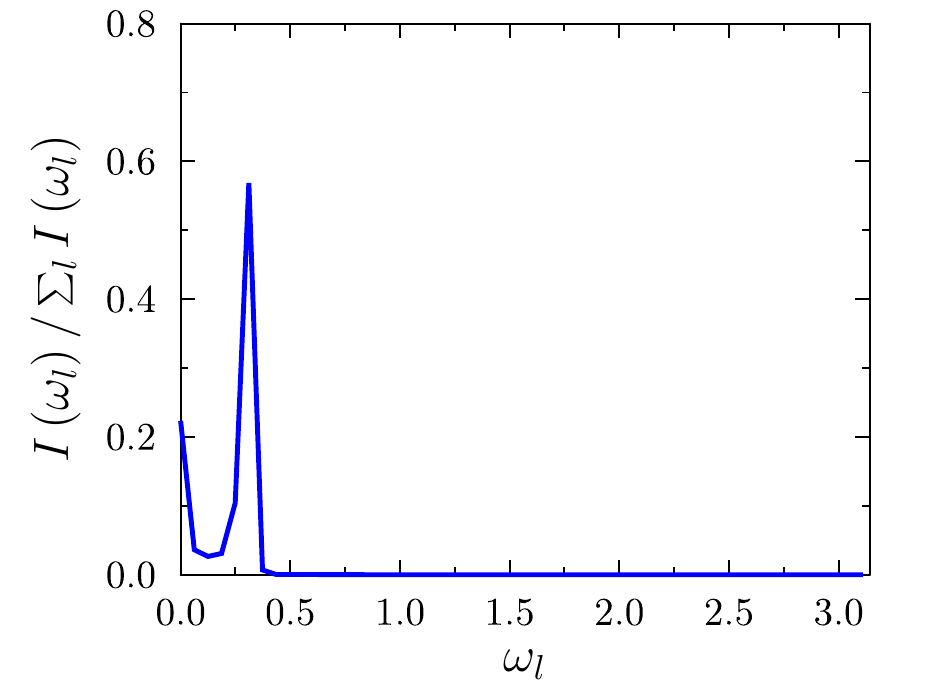}
\end{center}
\caption{(Color online) Normalized periodogram  $I(\omega)$ for $\langle N\rangle C_j$ calculated from ALICE data for $pp$ at $7$ TeV and $|\eta|<3$ \cite{ALICE}.}
\label{Fig-errors-1}
\end{figure}

A detailed discussion of the sensitivity of the modified combinants $C_j$ to the measurement uncertainties is given in \cite{Zborovsky}. Here we present only some remarks on the estimation of the errors in $C_j$ based on the published data and Monte Carlo simulations.

To summarize, as shown in Fig. \ref{Fig-errors}, one observes that statistical errors cause only some chaotic spread of the measured $C_j$ but do not result in periodic oscillations. In the case of monotonic behavior of $C_j$ as function of the rank $j$ (for example, when one uses the NBD) one gets no oscillations from errors. However, in the case when one observes oscillations,  systematic errors can actually blur the whole picture of oscillation (making them invisible). The most important point is that the oscillations of $C_j$ are highly correlated, cf. Eq. (\ref{Error}) (they are not chaotically scattered). Statistical errors do not give such oscillations.

Fig. \ref{Fig-errors}$(a)$  shows values of the $C_j$ for a NBD. Note that for large statistics (comparable to these in ALICE for $7$ TeV where we have $3.437\times 10^8$ selected events \cite{ALICE}) we obtain a very smooth dependence on the rank $j$. On the other hand, making analytical estimates of the size of the error one finds that
\begin{eqnarray}
Var\left[ \langle N\rangle C_j\right] &=& \left[ \frac{j + 1}{P(0)}\right]^2 Var\left[ P(j+1)\right] + \nonumber\\
+ &&\!\!\!\!\! \left( \frac{1}{P(0)}\right)^2 \sum_{i=0}^{j-1} \left( \langle N\rangle C_i\right)^2 Var\left[ P(j-i)\right] +\nonumber\\
+ && \!\!\!\!\! \sum_{i=0}^{j-1}\left[ \frac{P(j-i)}{P(0)} \right]^2 Var\left[ \langle N\rangle C_i\right]. \label{Error}
\end{eqnarray}
Note that the last term of Eq. (\ref{Error}) introduces dependence of the error in $C_j$ on the errors of all previous coefficients $C_{i<j}$. This results in an error cumulation effect leading to a significant increase of errors with increasing rank $j$, as can be observed in Fig. \ref{Fig-errors}$(b)$. However, despite such big errors, the values of $\langle N\rangle C_j$ lie practically on the curve (i.e., the points do not jump in the limits of errors). For such errors a Monte Carlo simulation would give the result presented in Fig. \ref{Fig-errors} $(c)$. Note that, as shown in Figs. \ref{Fig-errors} $(d)$ and $(e)$, using only statistical errors of the measured $P(N)$ results in a well defined curve. Finally, as demonstrated in Fig. \ref{Fig-errors} $(f)$, systematic errors provide limitations on the measured oscillations of $C_j$ in the form of some characteristic envelope (provided by the systematic uncertainties of $P(N)$, $P(N)\pm \delta[P(N)]$), around the mean values of $\langle N\rangle C_j$, which follows the fine structure of the oscillations of the residual mean values quite accurately.

We end with an estimation of the statistical significance of the oscillating behavior of the modified combinants $C_j$. This can be done using the periodogram-based Fisher $g$-statistic test \cite{Fisher,B-D}. This test determines whether a peak in the periodogram is significant or not and it proceeds as follows. Given a series $y(j) = \langle N\rangle C_j$  of length $L$, the periodogram $I(\omega)$ is first computed as
\begin{equation}
I(\omega) = \frac{1}{L} \bigg|\sum_{j=1}^{L} y(j)\exp(-i\omega j) \bigg|^2, \qquad \omega \in[0,\pi]. \label{P-1}
\end{equation}
It is evaluated at the discrete normalized frequencies
\begin{equation}
\omega_l = \frac{2\pi l}{L},\qquad l=0,1,\dots,a \label{P-2}
\end{equation}
where $a=[(L-1)/2]$  and $[x]$ denotes the integer part of $x$. If a  series has a significant sinusoidal component with frequency  $\omega_k$, then the periodogram will exhibit a peak at that frequency. Fisher derived an exact test of the significance of the spectral peak by introducing the Fisher $g$-statistic \cite{Fisher} defined as
\begin{equation}
g = \frac{\max I\left( \omega_l\right)}{\sum_{l=1}^{a}I(\left( \omega_l\right)}. \label{P-3}
\end{equation}
In Fisher's test, one is testing the null hypothesis, $H_0$, that the spectral peak is statistically insignificant against the alternative hypothesis, $H_1$,  that there is a periodic component in the signal $y(j)$. Under the Gaussian noise assumption, the exact distribution of the $g$-statistic under the null hypothesis $H_0$ is given by
\begin{equation}
P\left(g^{\star} > g\right) = \sum_{k=1}^{[1/g]} (-1)^{k-1} \frac{a!}{k!(a-k)!}(1 - kg)^{a-1}. \label{P-4}
\end{equation}

In Fig. \ref{Fig-errors-1} we present the normalized periodogram  $I(\omega)$ for $\langle N\rangle C_j$ calculated from ALICE data discussed here \cite{ALICE}. Note the large observed value of $g$, a peak in the periodogram, which indicates the existence of a strong periodic component and leads us to reject the null hypothesis. The probability that the spectral peak is statistically insignificant is $10^{-16}$.Therefore, similarly as in \cite{Zborovsky}, we conclude that notwithstanding the large sensitivity of the oscillations of the modified combinants to systematic uncertainties in the measurements of $P(N)$, they show enough power to disclose the fine details of experimentally measured multiplicity distributions, and can shed  new light on the dynamics of multiparticle production processes.

\end{document}